# Analyzing Problem Solving Using Math in Physics: Epistemological Framing via Warrants

## Thomas J. Bing[1] and Edward F. Redish[2]

[1]Department of Physics, Emory University, Atlanta, GA 30322, USA
[2]Department of Physics, University of Maryland, College Park, MD 20742, USA

**Abstract.** Developing expertise in physics entails learning to use mathematics effectively and efficiently as applied to the context of physical situations. Doing so involves coordinating a variety of concepts and skills including mathematical processing, computation, blending ancillary information with the math, and reading out physical implications from the math and vice versa. From videotaped observations of intermediate level students solving problems in groups, we note that students often "get stuck" using a limited group of skills or reasoning and fail to notice that a different set of tools (which they possess and know how to use effectively) could quickly and easily solve their problem. We refer to a student's perception/judgment of the kind of knowledge that is appropriate to bring to bear in a particular situation as *epistemological framing*. Although epistemological framing is often unstated (and even unconscious), in group problem solving situations students sometimes get into disagreements about how to progress. During these disagreements, they bring forth explicit reasons or *warrants* in support of their point of view. For the context of mathematics use in physics problem solving, we present a system for classifying physics students' warrants. This warrant analysis offers tangible evidence of their epistemological framing.

## I. INTRODUCTION

### A. Motivation

Mathematics is the backbone of physics. It provides a language for the concise expression and application of physical laws and relations. A student's development as a physicist entails, in no small part, becoming increasingly comfortable with mathematics. As physics teachers, we share a responsibility to help our students develop fluency with the mathematics of physics. But what does it mean "to become comfortable with mathematics in physics", how would we recognize it happening in a student, and how, as instructors, can we facilitate this process? In this article, we describe how we develop a clearer understanding of this issue by using videotaped ethnographic observations of groups of students solving physics problems in classes ranging from introductory algebra-based physics to graduate quantum mechanics.

What we learn is that although mathematics is an essential component of university level science, math in science is considerably more complex than the straightforward application of rules and calculation taught in math classes. Using math in science critically involves the blending of ancillary information with the math in a way that both changes the way that equations are interpreted and provides metacognitive support for recovery from errors. Our observations lead us to conjecture that expert problem solving in physics requires the development of the complex skill of mixing different classes of reasoning skills – the ability to blend physical, mathematical, and computational reasons for constructing and believing a result.

In order to analyze student behavior along this dimension, we have used the analytical tool of *epistemological framing*. This refers to the student's perception or judgment (unconscious or conscious) as to what class of tools and skills are appropriate to bring to bear in a particular context or situation. Although these framings are often tacit, in one particular situation they become much more explicit: when students argue with each other about what to do next. When there is a disagreement, the discourse tends to include *warrants* – explicit reasons for drawing a conclusion. We demonstrate that these warrants fall into clusters, casting a light on the way the student has framed the situation epistemically. This leads us to create system of classifying warrants that should provide a useful lens on the development of problem solving skills.

Our observations of students at a variety of levels of development of expertise leads us to suggest that, especially for students who are closer to the novice level, epistemological frames can be "sticky"; that is, students can get trapped in them for relatively long periods of time (many minutes).

These observations suggest that an epistemological framing analysis can be potentially useful for instruc-tion. First, it can suggest activities that might facilitate the development of expertise; and second, it can permit an instructor to recognize the development of sophisticated problem solving behavior even when the student makes mathematical errors.[1]

## B. An example

Many physics students struggle as they try to develop mathematical fluency. Part of their struggle is undoubtedly due to the sheer conceptual complexity of the mathematics commonly encountered in physics classes. As a physics major progresses through the undergraduate curriculum, she will encounter techniques requiring series expansions, three-dimensional vector calculus, linear algebra, complex numbers, differential equations, probability theory, and more. A robust understanding of any of these topics involves a complicated coordination of a large amount of information. University math departments commonly devote at least a semester-long class to each of the topics on that list.

This paper, however, focuses on a different aspect of the mathematical complexity our physics students encounter: epistemic complexity. "Epistemic" is an adjective referring to what one knows about the nature of (in this case, mathematical) knowledge. Simply put, the same piece of mathematics commonly fills many different roles in a physics class.

To illustrate epistemic complexity, consider the expression $x_f = x_i + <v>\Delta t$, relating an initial position $x_i$, a final position $x_f$, and average velocity, $<v>$, and a time interval $\Delta t$. First, this expression encodes a calculational scheme. If the object starts at $x_i = 3$ m and maintains an average velocity of $<v> = 4$ m/s for $\Delta t = 2$s, then the expression $x_f = x_i + <v>\Delta t$ tells you how to combine those given values to produce a numerical result for $x_f$: $x_f = 3$ m + (4 m/s)(2 s) = 11 m.

Second, $x_f = x_i + <v>\Delta t$, encodes a physical relation among measurements. An average velocity tells how far an object travels per given length of time. Multiplying by the time of the journey gives, $<v>\Delta t$, unwinding the definition of $<v>$ and representing how far you move in a given time interval. Tacking that on to where your object started from, $x_i$, must yield the position at the end of the journey, $x_f$.

Third, mathematics provides us with a concise system for recalling encoded rules and previously derived results. No one starts all physics problems from first principles every time. One can imagine a physicist simply quoting $x_f = x_i + <v>\Delta t$, and simply thinking to himself, "That's what the final position equation is."

Fourth, our sample expression $x_f = x_i + <v>\Delta t$ can be seen filling yet another role, highlighting another epistemic feature of mathematics in physics. It fits in with a large interrelated web of mathematical ideas. For example, it can be derived from the definition of average velocity by simple algebraic manipulation. It also has the conceptual structure of a base-plus-change symbolic form,[2] just like $v_f = v_i + <a>\Delta t$, and it coordinates with the interpretation of the determination of distance traveled from a velocity graph by the calculation of the area under the curve. Stepping even farther back, $x_f = x_i + <v>\Delta t$ can be seen as a solution of

$$\frac{d^2 x}{dt^2} = k \cdot$$

Mathematics thus fills many different epistemic rolls for a physicist. It reflects physical relations, provides a calculation framework, forms a web of interconnected ideas, and provides a packaging system for encoding rules and previous results. Even such an elementary expression as $x_f = x_i + <v>\Delta t$ displays this epistemic complexity. Developing expertise in physics includes learning how to coordinate these various natures of mathematics.

This paper discusses two detailed case studies of upper level undergraduate physics students at work on homework assignments in their physics classes. These students are grappling with the epistemic complexity of the mathematics, often struggling as they juggle various interpretations of the math at hand.

We have two main goals. First, we propose a cognitive mechanism that helps us model their thinking as they juggle these interpretations of their math (epistemological framing). Second, we offer a convenient, powerful way to gather evidence of this cognitive mechanism from students' speech (examine the warrants students use in their discourse). We hence turn first to overviews of the framing literature and argumentation theory.

In Sec. II we discuss the cognitive framework we are working in including giving an explanation of the concept of framing. We also give a brief review of the relevant elements of argumentation theory. In Sec. III we discuss our methodology. In Sec. IV we identify the four epistemic framings we find in our data, and in Sec. V we present our two detailed case studies. We present our conclusions and the implications for instruction and future research in Sec. VI.

## II. THEORETICAL BACKGROUND

We begin by situating ourselves within a particular theoretical way of thinking about student thinking: the Resource Framework (RF). We give a brief summary of the assumptions of this approach. We then provide a simple example ("Sarah shifts her reasoning") to illustrate how epistemological framing fits within the RF. We then discuss framing in general and epistemological framing in particular.

## A. The Resource Framework: A brief overview

There are many theoretical lenses that are available for building models of student behavior. The one we use is the Resource Framework (RF) that has evolved out of the "Knowledge in Pieces" approach developed by diSessa and his collaborators.[3][4] This framework is documented in detail in a number of published papers[5][6][7][8] and in an evolving Wiki at the Physics Digital Library PER-Central website.[9] We present a brief overview here but encourage readers interested in more detail to access the original papers and the website.

The Resource Framework (RF) is a structure for creating phenomenological models of high-level thinking. It is based on a combination of core results selected from educational research phenomenology, cognitive/neuroscience, and behavioral science. It is a framework rather than a theory in that it provides ontologies – classes of structural elements and the way they behave – and it permits a range of possible structures and interactions built from these elements. As such, it provides a framework that permits the creation of descriptive and phenomenological models that bridge many existing models such as the alternative conceptions theory and the knowledge in pieces approach or cognitive modeling with the socio-cultural approach. The RF does not (as yet) create mathematical models in which predictions arise from calculations.

The RF is an associative network model with control structure and dynamic binding.

*1. Network* – The basic ontology of the RF is that of a network, built from the well-established metaphor of neurons in the brain. The activation of knowledge is thought of as the activation of clusters of linked neurons.

*2. Associative* – The activation of one resource or cluster of resources leads to activation of other clusters. Learning is pictured at a fundamental level as the establishment of strong connections so that activation of one resource or cluster of resources leads inevitably to the activation of other resources. Associations can be excitatory (encouraging activation) or inhibitory (discouraging activation).

*3. Control structure* – The network of resources in the brain is not simply associative. There brain has structures (hippocampus, cingulate gyrus, etc., etc.) that appear to have specific purposes, just as in other parts of the body (heart, lungs, liver, etc., etc.). A major structure that has particular relevance for us is the prefrontal cortex, which is where perceptual information is mixed with information in long-term memory to prime appropriate responses and actions. The evaluation of a perceived situation affecting action is a very well documented component of behavior in mammals in situations ranging from simple conditioning of rats to complex social behavior in humans. (See for example Fig. 1 and the extensive references in Redish, Jensen, and Johnson.)[10] Control structures rely heavily not only on activating association, but also on inhibition.

*4. Binding* – Clusters of resources that activate together frequently become strongly tied so that they always activate together.[11] This makes possible the creation of networks of higher level structures – concepts, p-prims, or schemas – that the user perceives as unitary. Binding can occur at many levels from being extremely tight (e.g., it is hard to see the word "cat" and not to imagine its referent) to being rather loose (e.g., one can perceive an orchestral performance as unitary or listen for individual instruments or motifs). Both basic associations (concepts) and control structures (framings) can be tightly or loosely bound.

*5. Dynamic* – A critical element of the entire model is that it is extremely dynamic. Associations are activated and inhibited depending on context.

In this paper we are building a component of the control structure appropriate for using mathematics in physics: epistemological framing.

## B. A framing story: Sarah shifts her reasoning

An analysis in terms of epistemological framing focuses on the moment-to-moment shifts observed in students' reasoning. Their interpretation of the task and knowledge at hand can change, as in the following example.[12] "Sarah" is an upper-division undergraduate physics major who sat for an interview aimed at her understanding of electrical conductors and insulators. Sarah has just explained how insulators are so dense that current cannot flow through them. Wanting to explore this further, the interviewer brings up the case of Styrofoam. When the interviewer asks her whether Styrofoam is an insulator, Sarah responds that it is. Her response to the interviewer's question, "Why?" is that she "memorized it". The conversation continues, and when the next opportunity arises for Sarah to justify a claim she makes a blanket statement citing "organic chemistry". So far in the interview, she is relying on authority in her explanations, quoting rules and facts.

After the interviewer prods her to give "any explanation you find," Sarah's reasoning undergoes a shift. She gives a more detailed, more conceptual account of conductance. Sarah puts together a little story about electrons getting torn away from their parent atoms and then being free to move. She explains how a battery could perhaps cause this electron-tearing and how a higher temperature wire might also have more energy available to tear electrons off the atoms.

The shift we care about in Sarah's reasoning concerns the types of explanations she gives. She began by quoting facts. Implicit was Sarah's epistemic interpretation of her situation and the interviewer's intentions. What is the nature of the knowledge in play here? "Oh, OK, this interviewer wants factual information about conductors and insulators. I'll give him some facts I remember."

The interviewer's apparent dissatisfaction with her quoted facts and subsequent "any explanation you find" prompt caused Sarah to reinterpret the activity. She came to see the interviewer's questions as prompts to tell a story about conduction. Sarah is less sure of her story about tearing off electrons than she was about her quoted facts, but she sees this uncertainty as permissible now. Epistemically, "now we're constructing stories, not quoting facts."

Briefly, Sarah has a different epistemological framing of her activity in the two parts of this episode. The different epistemological framings, different implicit answers to "What kind of knowledge is in play here?" led Sarah to bring different subsets of her knowledge store to bear on the interviewer's questions.

We now turn to a more detailed account of this epistemological framing process, beginning with an overview of epistemic resources.

## C. Epistemological resources activate and deactivate in Sarah's reasoning

In our brief example Sarah treats knowledge as two different types of things. She begins by viewing knowledge as fact-based and authority-driven, later shifting to seeing knowledge as a personally constructed thing. We describe this shift by saying that different epistemological resources have been cued in Sarah's mind.

An epistemological resource is a cognitive modeling element. It represents a tightly bundled packet of information that, when activated by the mind, leads the individual to interpret the knowledge at hand in a certain light. But an epistemological resource is a control structure, not a concept; epistemological resources affect how students perceive the nature of the situation under current consideration and they control what conceptual resources are brought to bear. Do they see scientific knowledge as fixed and absolute or as being relative to one's point of view? Do they view scientific knowledge as something they can construct for themselves or as something that must be handed down from an authority figure?[13][14]

Epistemological resources, like other resources, are dynamic; they can activate and deactivate during the moment-to-moment flow of an episode. Broad, decontextualized questions such as "Do you see science knowledge as being handed down from authority?" at least by themselves, are unlikely to elicit meaningful information on students' functional epistemologies. Such a question assumes that students have relatively stable, context-independent beliefs about the nature of science. Much like the case with conceptual knowledge,[3][6] authors have argued that students' epistemic stances are manifold and highly sensitive to context.[15][16] Sarah, for example, displayed a shift from "knowledge as authority driven" to "knowledge as constructed by oneself" in her brief electric conduction interview. This shift happened in response to an interviewer's prod. It was an in-the-moment reaction to the natural flow of the conversation. One would certainly not expect that this isolated shift signals a large-scale change in Sarah's approach towards physics. It is unreasonable to think she never saw physics as being about telling conceptual stories before, nor is it reasonable to think she will never quote authority again. There are many similar published examples of in-the-moment shifts in students' reasoning.[6][7][17][18]

As further evidence of the manifold nature of students' epistemologies, there also tends to be a disconnect between how students view the nature of formal science and how they proceed to interpret their own work in science class.[19] Epistemological stances evolve, in a time averaged sense, in complex ways as a student progresses through his education.[20]

On more local timescales, many epistemological resources are available to students. These epistemological resources are often closely correlated with certain bits of the wide range of conceptual information available to students. Sarah's activation of "knowledge as authority-driven" pointed her towards her store of organic chemistry facts (or perhaps vice-versa).

With a wide range of conceptual and epistemological possibilities available, a model of students' thinking must also include a process by which the set of all possible epistemological and conceptual options is pared down to a manageable size for conscious consideration by the individual. That process is called *framing*. This paper proposes a specific epistemic lens (looking at the warrants used in their arguments) for analyzing how upper level physics students are framing their use of mathematics.

## D. Framing:
## What kind of activity is going on here?

So far, we have been using the term "framing" in what amounts to a "common speech" mode. We have not defined it explicitly. But in the behavioral sciences, especially anthropology and sociolinguistics, "framing" represents a specific technical term. *Framing* is the, usually subconscious, choice the mind makes answering the question, "What kind of activity is going on here?" It narrows down the set of all possible mental options to a manageable subset. An individual's framing of a situation tells them what is necessary to pay careful attention to in a situation and what can be considered irrelevant and ignored. This "selective attention" reduces processing load and is the benefit created by the mental structure that permits framing.[21] Framing is a common, everyday cognitive process.

As a quick example, consider entering a hotel. Even if you have never been in that particular hotel before, you will immediately have a general idea how to proceed. You would expect there to be a front desk with a check-in clerk, lots of numbered rooms organized in a particular way, and perhaps a restaurant or two. You would plan on doing certain things in this building like sleeping and preparing for the next day's business. You would also have social expectations. You would not plan on shouting across the lobby, playing your television at full volume at 11pm, or throwing furniture off the balcony.

Framing should not be equated to activating a large, stable instruction list. It's not as if you immediately run down a checklist upon entering a hotel. Where's the elevator? There it is. Where is the concierge desk? There it is. Where is the restaurant? There it is. Large data structures like this hotel list are like a set empty slots ready to be filled in with the particulars of a situation. Several early studies in artificial intelligence (from which modern framing studies partially evolved) were concerned with identifying (and then programming) such data structure "frames".[22][23][24]

This paper does not equate framing with the recall and activation of organized, rigid data structures. Rather, we use framing as the cuing of fuzzy, adaptable networks of cognitive resources. Not finding a restaurant in your hotel doesn't necessarily destroy your interpretation of your surroundings as a hotel. If the room numbers' organization isn't the standard floor-by-floor numeric order, you would likely still be able to find your room eventually.

Framing has been studied in a wide array of academic disciplines including linguistics, sociology, art, psychology, and anthropology.[25][26][27][28] All of these studies implicitly agree on the existence of what has been called "Felicity's Condition".[29] Felicity's Condition is the unspoken premise naturally adopted by an individual that incoming information, whether it be spoken, read, observed, etc, comes from a rational source, and it is thus up to the individual to attempt to contextualize and hence interpret that incoming information. Framing is the process by which the mind attempts this contextualization and interpretation.

Different individuals can certainly frame the same incoming information in different ways.[27] A quick example is to note that what may be play to a golfer is work to the caddy.[30] Miscommunications can arise when two individuals frame their interaction differently, each bringing a different subset of their available resources to bear on the situation. Framing should not be thought of as something that happens only once at the start of a new activity. People continually recheck their framing of a situation and may alter it accordingly, bringing new resources into conscious consideration while temporarily disregarding other ones.[31] Sarah and her discussion of electrical insulators is one such example.

Framing can lead people to subconsciously disregard some strands of input information that are not seen as currently relevant. A latecomer taking his seat at a theater can be ignored, possibly not even noticed, by other audience members.[26] The students in this paper's later case studies display an analogous selective attention. They can seem temporarily oblivious to a mathematical course of action that may be obvious to a classmate, instructor, or researcher. This selective attention is a direct result of their epistemological framing of their mathematics. They, in that moment, are interpreting the math at hand in a certain way, focusing on a particular aspect of the math knowledge in play.

## E. Argumentation theory helps us
## get evidence of framing

Framing is often unconscious, even unnoticed by the person doing it. How, then, can a researcher gather evidence of how these upper level physics students are framing their use of mathematics? How can we identify what they see as "the particular nature of the math knowledge in play"? A possible solution was suggested in our observations of students working on physics problems in groups. When students disagree on a procedure or result, they are often explicit about *why* something should be believed or not. These comments, referred to as *warrants* in argumentation theory, show us what epistemic assumptions they are making at the moment. We therefore suggest: look at the warrants they use in their math arguments. We hence turn to a brief overview of argumentation theory.

There are several subfields that are sometimes colloquially lumped under the rubric, "argumentation theory".[32] On one end of the continuum is what is best called formal logic. Studies in formal logic deal with relatively clean and straightforward methods of proof-making that can easily be decontextualized from whatever given situation is at hand. The formal logic chains that result from such analysis, chains like "If A then B, if B then C but not D, etc," lend themselves readily to computational modeling,[33] although even such apparently straightforward applications of classical logic rely on fuzzy mental processes that are very difficult to describe in detail analytically.[34]

A second branch of research, the one that is most often actually called "argumentation theory", includes what is often called rhetoric.[32] This field of research focuses most on presenting, as opposed to having, an argument. It attempts to parse the content of a given argument into some kind of structure and often carries some sort of evaluative tone with regard to that structure. A central pillar of this field, and an important basis for this paper's analysis, is the work of Stephen Toulmin. He devised an often-cited system for parsing an argument into such parts as *claims*, *data*, and *warrants*.[35] A person will make a statement, the claim, that requires proof. They will then offer one or more relevant facts, the data. The warrant is the bridge, sometimes unspoken, that explains how the given data relates to the claim at hand. For example, I might state that Thomas Jefferson is the greatest American founding father (claim) because he largely wrote the Declaration of Independence (data). The relevant warrant that would link this data to that claim would be that the Declaration of Independence is a cornerstone document of the United States, laying out the nascent country's case for autonomy.

Because argumentation theory deals more with real-world arguments than formal logic, analysis schemes like Toulmin structures are best thought of as heuristic guides for parsing arguments, not formal organizers. Attempts to carefully map out even the structure of a published, formal legal argument according to Toulmin's scheme resulted in an explosion of complexity.[36] Researchers found it increasingly necessary to add sublevel after sublevel to the basic claim-data-warrant scheme as they encountered more and more interwoven lines of reasoning. Even with all these complicated sublevels in the argument's diagram, they had trouble accounting for large chunks of implicit information that the writers of the legal document simply assumed the reader would know (cf. Ref 34). Another study trained a group of corporate professionals in Toulmin structures and then had them try to apply this tool to diagram an argument relevant to their profession. Their success was limited, and many participants noted that the resulting argument diagrams were less convincing than the original arguments.[37]

We see these arguments as pushing a Toulmin analysis too far; somewhat analogous to Hilbert's or Russell and Whitehead's attempt to create a complete axiomatic structure for mathematics. It is now known that such a complete structure is impossible even for mathematics, and the interaction of a student with a physics problem is not in fact formal mathematics. It is more closely akin to natural discourse.

Naturally occurring arguments are more nebulous than an argument fitting a clean Toulmin structure. Justifications that may be logically unsound can be cognitively sound – completely acceptable and compelling in informal, real-time situations according to complicated, probabilistic mental processes.[38] A branch of research, often gathered under the label "discourse analysis",[32] concerns itself primarily with the in-the-moment patterns people employ in their speech and thought as they construct and communicate arguments. These in-the-moment argument constructions are often verbally incomplete. They often refer to a body of knowledge that the speaker (correctly or incorrectly) assumes he shares with the listener. These flow-of-conversation arguments sometimes have holes in them that are consciously or unconsciously overlooked.

This paper's work will most closely align with this discourse analysis research approach. It takes a detailed look at physics students' mathematical arguments, but it does not attempt to analyze these arguments according to a formal, computational structure of logic. As later examples demonstrate, these students' thinking is too dynamic to allow such a structural interpretation. This paper merely borrows Toulmin's idea of a "warrant" to help analyze a vitally important "in-the-moment pattern" in physics students' speech and thought: their epistemological framing of the math at hand. Students' warrants can shift from moment to moment, and these shifts are closely tied to what they interpret as the nature of the math knowledge currently in play.

Practically speaking, this shifting of warrants results in physics students giving different kinds of proof at different times during a mathematical argument. The analysis of the students' mathematical warrants offers a powerful window to describing how they are currently framing their activity.

The idea of different kinds of proof being accepted in an argument is not, in general, a new one. On a grander scale, researchers have noted that what counts as valid proof does not necessarily remain the same as one crosses social or cultural boundaries. One needs look no further than the Creationist/Evolution debate for an example.[39] On a smaller classroom scale, this phenomenon of shifting justification has also been noted with biology students.[40]

The idea of different kinds of reasoning counting as sufficient proof has also been noted in mathematics education research. Researchers have discussed, for example, the *embodied*, *proceptual*, and *formal* reasons $13 + 24$ equals $24 + 13$.[41 42] The embodied explanation is that adding twenty-four objects to a collection of thirteen objects gives you the same total number as if you started with thirteen objects and added twenty-four. A proceptual explanation focuses on how you can manipulate the meaning-laden symbols in the problem in a prescribed manner, i.e. you can do the column-addition you learned in elementary school, and get the same result either way. The formal reason $13 + 24$ equals $24 + 13$ is that it's assumed true by axiom. It's the commutative property.

Two extended case studies demonstrate how the flow of a physics problem solving conversation can be parsed by viewing it as two or more individuals trying to juggle and coordinate various types of mathematical justifications – epistemological framings – in their reasoning.

## F.  A summary note on terminology

In summary, we use three terms to describe our epistemological control structure: epistemological framing, epistemological resources, and warrants.  Note that by introducing these distinct terms, we are not proposing that these correspond to three distinct cognitive structures.  Although that may be the case, we do not at this time have evidence to support that claim.  Rather, we use these three terms to provide a way of emphasizing different aspects of what may, in the end, turn out to be a reasonably unitary and non-separable process: the process of making a judgment about what knowledge applies in a particular situation.

The use of the term *framing* focuses our attention on the interaction between the cue and the response.  It stresses that there is an evaluation based on previous general knowledge and experience that is taking place. This evaluation is very often a subconscious one. This paper's explicit focus on the warrants observed in physics students' mathematical arguments is precisely an attempt to define explicit evidence for the inherently implicit cognitive process of framing.

The use of the term *resource* focuses our attention on the fact that the kinds of reasons students cite fall into broad categories with a common underlying structure.  The use of the term *warrant* focuses our attention on the epistemic nature of the specific argument being made.  For example, a student pointing to a five-line calculation he just performed and a student pointing to an electronic calculator's output may be using slightly different specific warrants (i.e. you can trust a careful manual computation vs. you can trust a machine's algorithmic churning).  Both students' reasoning, however, falls under the same general class of warrant: algorithmically following a set of established computational steps should lead to a trustable result.  We would call this general class of warrant an epistemic resource.  It's a general, repeatedly observable way students view knowledge.  This epistemic resource acts as a cognitive control structure.  It is one (of many) regulators of students' epistemic framing (i.e. how they implicitly choose to interpret the knowledge at hand).

In the case studies that follow, we are somewhat loose with our specific uses of "class of warrant" vs. speaking of one particular flavor of warrant specific to the situation at hand.  There is simply not always sufficient evidence to carefully distinguish between the two, nor is it vital to our general argument (that focusing on warrants gives explicit evidence of epistemic framing) to do so.

## III. METHODOLOGY

Before turning to our two extended case studies, we give a brief description of the video data used in this study and how our common framings emerged from this data set.

Approximately 150 hours of raw video data of upper-level undergraduate physics students was collected for this study.  These students were enrolled in such classes as Quantum Mechanics I and II, Intermediate Mechanics, Intermediate Electricity and Magnetism, and Intermediate Theoretical Methods.  Most were physics majors.  None were in their first university physics class.

About 80 of these 150 hours (including both this paper's case studies) come from group homework sessions.  Our students would routinely meet outside of class to work on their homework together.  We simply would ask their permission at the start of the semester to video tape these meetings.  Another 25 hours of the video data set came from individual problem-solving interviews with students. The rest of the video set was made of actual classroom recordings. These classroom videos tended to be less useful for this study since they contained a much smaller amount of student speech.

## A. Students' framings are easiest to identify via contrasts and shifts

Evidence for how students are framing their math use is easiest to pick up when there is some sort of contrast or misunderstanding present. Such framing confusions are common sources of disagreements, even in non-physics settings.[43] Many mathematical disagreements physics students have with each other reduce to the first student essentially saying "Look at this math issue this way" while the second student is claiming "No, you should be looking at it this other way". The students are debating which aspects of their mathematical knowledge are currently relevant. Examining the warrants[35] physics students use in their mathematical arguments offers a good window to how they are currently framing their math use

## B. Data selection process

We needed some sort of selection process that could pare down our 150-hour data set to a collection appropriate for a close, careful analysis. The first author was present during 95% of the tapings themselves and took detailed notes of the students' activity. These notes allowed the video databank to be quickly searched for the best debates, arguments, and misunderstandings. At this early point in the pare-down process, "best" simply meant the debates and arguments that most likely had a lot of material available for possible analysis. Sometimes "best" translated to a simple clock reading. If students spent five minutes arguing about a certain point, there was a good chance a closer analysis might find a relatively large amount of speech that clearly annunciates their ideas. Other times the "best" arguments were selected for the novelty of their content. An argument about whether an expression simplifies to $x^2 + 2x + 1$ or to $x^2 - 2x + 1$ is likely to be routine. The students are likely to quickly agree on a useful way to resolve the argument. They are likely to share a common framing, which means there won't be much explicit evidence for that framing. However, an argument about a novel issue is much more likely to bring about a variety of approaches, a variety of framings.

The first pass through the 150-hour data set yielded about 50 snippets containing the arguments, debates, and misunderstandings most likely to be explicit and long enough to offer good evidence (i.e., clearly identifiable mathematical warrants) for how the students were framing their math use. Eventually, a framing analysis was carried out on other episodes that didn't contain such obvious arguments. Such an extension helped to assure the generality of our framing analysis. The reader is invited to look at the dissertation from which this paper is drawn for such non-argument examples.[1b]

Our 50-snippet subset of arguments was meant to offer the best evidence for deciding whether a set of common framings exist and, if they do, what they specifically are. The next section describes the methodology used to help these common framings emerge.

## C. Knowledge analysis: Common framings emerge from the data set

In order to make sense of our data, we performed a knowledge analysis.[2][44] The basic idea is to find a common thread to condense the episodes according to a common analysis scheme. Knowledge analysis is an iterative methodology.

Specifically, this project began with identifying a small sample of about 50 episodes that were most likely to contain relatively easy-to-spot frame shifts. A subset of these 50 sample episodes were analyzed individually at first, the goal being to describe what type of warrants were the students using in their mathematical arguments.

Once a small collection of these individual analyses were collected, it became possible to look for consistencies across episodes. Several clusters incorporating similar individual math framing examples were identified. The next step was to do a similar analysis on a new set drawn from those 50 episodes and see if these original clusters could incorporate these new examples of students' mathematical thinking as well. Appropriate changes were made to the clusterings in light of this new data set, and then a third set of episodes were considered. After several iterations, the clustering scheme stopped evolving significantly. Eventually the whole 50 episode subset was used, with each individual episode cycled through more than once.

Four main clusters emerged from this data set's examples of physics students' framing of their math use. They capture four general types of justification these students offer for their mathematics: "Calculation", "Physical Mapping", "Invoking Authority", and "Math Consistency". These clusterings are not meant to be mutually exclusive or sufficient to span all possibilities. They are merely presented as the most convenient way found of structuring comparisons across many different episodes in our data set.

## IV. THE FRAMING CLUSTERS THAT EMERGED FROM OUR DATA SET

Our knowledge analysis led us to classify the student interactions into four common framing clusters: Calculation, Physical Mapping, Invoking Authority, and Math Consistency. These framings parallel the discussion of $x_f = x_i + <v> \Delta t$ in the introduction. We begin with a brief overview of each and then present a more complete discussion and comparison of the four clusters. Finally, we discuss inter-rater reliability.

### A. Framing 1: Calculation

A calculation framing, like all the other framings that emerged from the data set, is primarily identified by the general class of warrant students choose to use. In this case the epistemological resource (i.e. the general class of warrant observed) is: *algorithmically following a set of established computational steps should lead to a trustable result*. The specific warrants used, like all the other warrants we identified in our data set, couples closely to the epistemological resources currently activated by the student. Epistemological resources, recall, are control structures. They lead the student to frame the knowledge at hand in a certain way, which focuses the student's attention on a particular subset of his total knowledge.

In a calculation framing, students rely on computational correctness. The warrant may be implicit, especially in non-argumentative settings. If an instructor were deriving $y_f = y_i + v_i t - \frac{1}{2} g t^2$ from $\frac{d^2 y}{dt^2} = -g$, she would probably just carefully explain her steps to her students. They would likely accept the result without further thought. It is rare to explicitly explain, "OK, because carefully following a set of computational steps allows one to trust a result, we should trust this derivation." It would rely on an unspoken epistemological resource, one that's shared because both instructor and student frame the discussion as calculation.

### B. Framing 2: Physical mapping

When physics students frame their math use as physical mapping, they support their arguments by pointing to the quality of fit between their mathematics and their intuition about the physical or geometrical situation at hand. This class of warrant can be associated with the epistemological resource: *a mathematical symbolic representation faithfully characterizes some feature of the physical or geometric system it is intended to represent*. Again, it is through identifying these (relatively explicit) warrants that a researcher can get information about the (relatively implicit) epistemological framing process in the student's mind.

For example, suppose we wanted to explain why the expression for the force exerted by a spring, $F = -kx$, includes a negative. We might explain how stretching a spring makes it pull backwards as it tries to contract back to its natural length. If you compress the spring, it'll push back against the compression as it tries to expand. In both cases the spring force is opposite the way the spring is deformed. That is, if $kx$ is positive (say you extend the spring to the right) then the spring pulls in the negative (i.e. left) direction. If $kx$ is negative (say you compress the spring leftwards) the spring exerts a force to the right (positive) direction. Again, we do not necessarily have to spell out an explicit warrant or our math-should-model-the-world epistemological resource. They come along with a physical mapping framing.

There is a more general point about distinguishing a calculation framing from a physical mapping framing. At some level, all mathematics is ultimately grounded in physical experience. A child learns to associate "1" with a single object, "2" with a collection of two objects, and so on. Higher and higher mathematics are built up by analogy and extension of what are ultimately physically grounded ideas.[45][46] The distinction between a calculation framing and a physical mapping framing largely concerns a person's in-the-moment awareness of the physical referents of her math.

We note that for this work, we do not distinguish between the use of physical statements from the use of abstract geometrical statements as warrants. This is because in the examples we observed, the geometry arose out of the location of the physical situation in 3-space. We expect that if a wider class of situations were considered, it might be appropriate to separate physical and geometrical framings.

### C. Framing 3: Invoking authority

Suppose we were trying to convince you what the rotational inertia of a solid sphere was. We might simply pick up an introductory physics book, thumb through the index until we found "rotational inertia", turn to page 253, and point at an entry in a table that says "solid sphere, $I = \frac{2}{5} MR^2$". Perhaps you would accept our argument, also accepting the implicit class of warrants (i.e. epistemological resource) that underlies our reasoning: *information that comes from an authoritative source can be trusted*.

An invoking authority framing is often closely tied to finding the right level of detail to go into during a problem. It is unreasonable to take every single prob-lem down to absolute first principles every time. Some results will always simply be taken for granted. Perhaps you would be more likely to accept our earlier argument for the rotational inertia of a solid sphere if we were engaged with a larger problem like finding the time it would take such a sphere to roll down a given ramp. You might judge the specific value of the sphere's rotational inertia to be sufficiently irrelevant to the problem's main purpose to permit us to quote from the textbook.

Another common trait of the invoking authority framing is the absence of extended chains of mathematical reasoning. "Chaining" has been closely tied to students' mechanistic reasoning.[47][48] When a student links together a series of implications, she is chaining. An example would be "adding another resistor in series puts another obstacle in the current's way, so the total resistance goes up, but the battery's push remains the same, so the current flowing decreases". Students engage in mathematical chaining arguments as well. The calculation framing often cues reasoning like "$A = BC$, but we don't know $C$, but we can use $C = EF$ to get $C$, then we can use $C$ to get $A$". The electric current example just above could be a nice example of chaining while in a physical mapping framing if the student was simultaneously thinking about the formula $\Delta V = IR$. Chaining is usually absent or severely limited if the student is framing his math use solely as invoking authority.

## D. Framing 4: Mathematical consistency

Suppose you were trying to explain Coulomb's Law for the electric force, $\overrightarrow{F_e} = \dfrac{1}{4\pi\varepsilon_o}\dfrac{q_1 q_2}{r^2}\hat{r}$, to a student. You might remind him of the expression for the gravi-tational force, $\overrightarrow{F_g} = -\dfrac{Gm_1 m_2}{r^2}\hat{r}$, and demonstrate how ideas from this more familiar bit of math map to Coulomb's Law. Both forces depend on the relative strengths (mass or charge) of the two objects in question. Both forces fall off with respect to distance in the same way, and both include a proportionality constant ($G$ or $\dfrac{1}{4\pi\varepsilon_o}$) that must be experimentally measured.

Even dis-analogous observations can be illuminating. Gravity is always attractive, hence the negative sign is explicitly included in front of the always positive masses. An electric force can be attractive or repulsive, so the implicit signs on the positive or negative charges, $q_1$ and $q_2$, will determine the direction of the Coulomb force.

Implicit in your discussion with the student would be the class of warrants indicative of a Math Consistency framing: *mathematics and mathematical manipulations have a regularity and reliability and are consistent across different situations.* Establishing a common underlying mathematical structure allows one to trust the relevant set of relations and inferences.

## E. Correlates of the four framing clusters

Our four common framings are primarily identified via the warrants physics students use in their mathematical reasoning. Other indicators, however, have been observed to cluster preferentially around certain framings. The table below summarizes these primary (i.e. warrants) and secondary framing indicators we have observed in our data set.

| | **Calculation** | **Physical Mapping** | **Invoking Authority** | **Math Consistency** |
|---|---|---|---|---|
| **Class of Warrant Used** | Correctly following algorithmic steps gives trustable result | Goodness-of-fit between math and physical observations or expectations attests to a result. | Authoritatively asserting a result or a rule gives it credence. | Similarity or logical connection to another math idea offers validation. |
| **Other Common Indicators** | -focus on technical correctness<br>-math chaining: need this to get that | -often aided by a diagram<br>-demonstrative gesturing | -quoting a rule<br>-absence of mechanistic chaining<br>-little acknowledgment of substructure | -analogy with another math idea<br>-categorization |

*Table 1: Four common framings and their primary (i.e. warrants) and secondary indicators.*

Framing is a dynamic cognitive process. A person's mind makes an initial judgment regarding the nature of the situation at hand, but that judgment is continually updated and reevaluated. New information comes to the student all the time, whether in the form of a classmate's comment, an interviewer's interjection, simply turning to a different page in a textbook, or even spontaneous random associations within her own brain. This new information can lead a student to reframe her activity. As a result, the epistemological framings observed in these students' work can extend over a range of time periods. We have found examples in our data set ranging from ten seconds to ten minues.

## F. Inter-rater reliability of epistemological framing analysis

The value of this study's epistemological framing analysis depends in part on how readily other researchers can apply it consistently. An inter-rater reliability study was carried out by giving this paper's methodology discussion to another researcher and that researcher was then asked to parse a new transcript for epistemological framing. Details of this inter-rater reliability test are given in Chapter Four of the first author's dissertation.[1b] Different researchers agreed on their framing codes 70% of the time for a novel transcript before any consultation or discussion. This figure improved to 80% after discussion.

We do not expect our epistemological framing coding scheme to yield a 100% consistent coding of a random transcript. Students' thinking is simply not that cleanly compartmentalized. Indeed, we argue (see ref. 1b, Chapter 7) that one characteristic of expertise in physics problem solving is the ability to effectively blend these four framings dynamically.

Two issues are relevant here. First, there is the question of how often students are observed to spend an appreciable time, say a minute or more, uniquely in one of this paper's four common framings. Of all the data analyzed for this study, perhaps less than 50% can be cleanly coded in minute-or-longer chunks under one of these general framings.

The second notion of "clean coding" of framing that is relevant concerns not these minute-long pure state framings but rather our ability to identify smaller chunks in hybrid framings. Calculation, physical mapping, invoking authority, and math consistency do a reasonable job of spanning the space of these students' mathematical arguments. We observed that about 90% of a random episode or more can be seen as made up of behavior indicative of those four landmark framings. But at this stage of student development (upper division physics majors) hybrids are common.

Perhaps a student quotes a few computational rules as he performs a long calculation. Maybe a student makes an analogy to both a similar physical situation and a similar math structure. As the inter-rater reliability test shows, researchers can still use this study's analysis scheme to identify evidence of these elemental framings in a piece of transcript that is, in general, a more complicated (and a more expert-like) hybrid framing.

## V. CASE STUDIES

We now turn to our two case studies. These detailed examples illustrate how this paper's warrant-based framing analysis can be applied to parse an authentic conversation among physics students.

The students' framing of their math use plays a significant role in each of these episodes. The principal dynamic in each of these conversations concerns how to interpret the math at hand. A significant amount of these students' energy goes into trying to establish the epistemological framing they see as appropriate.

In both studies, their thinking is dynamic. Different bits of their mathematical knowledge are activated and deactivated as they frame and reframe their activity. Sometimes framing differences have marked effects. The students sometimes talk past each other, neither one seeming to hear what the other is saying, because they are framing their work differently. Sometimes a student's framing can exhibit considerable resistance to change, as in this section's first case study. The second case study shows students being more flexible in their framing.

## A. Case study 1: Framings can have inertia

The first case study comes from a group of three students enrolled in the class Intermediate Theoretical Methods (PHYS 374). One is a junior (S2) and the other two are sophomores (S1 and S3). These three students met regularly outside of class to work on their homework together, and this episode was taped during one such homework session.

### *1. The question*

Our episode starts in the middle of their work on one of that week's homework problems. The problem they were considering reads:

A rocket (mass *m*) is taken from a point A near an asteroid (mass *M*) to another point B. We will consider two (unrealistic) paths as shown in the figure. Calculate the work done by the asteroid on the rocket along each path. Use the full form of Newton's Universal Law of Gravitation (not the flat earth approximation

"mg"). Calculate the work done by using the fundamental definition of work: $W_{A \to B} = \int\limits_{A}^{B} \vec{F} \cdot d\vec{r}$ .

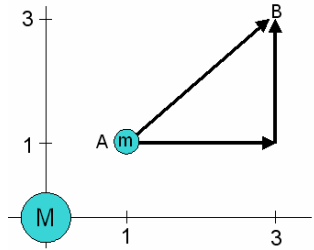

*Figure 1:  Case Study 1 problem*

The reader familiar with the physics of this example will recognize an attempt to get the student to see how the theorem that says potential energy is path independent arises out of explicit calculation.  Not all of the students in this discussion recognized the point of the problem from the beginning.

### *2.  The first framing clash*

During this episode, the students are trying to decide if the work done should be the same along the two paths from A to B. They had previously suppressed the $G$, $m$, and $M$ constants and written the (incorrect)[49] equation $\int\limits_{\sqrt{2}}^{3\sqrt{2}} \frac{1}{r^2}\,dr = \int\limits_{1}^{3} \frac{1}{y^2+9}\,dy + \int\limits_{1}^{3} \frac{1}{x^2+1}\,dx$ on the blackboard to express the work done along the direct and two-part paths, respectively. They have also copied the diagram of the situation from the problem statement.

The students are standing at the blackboard where all the relevant equations and diagrams appear. We focus on the type of justification each student offers for his math arguments:

1. **S1**: what's the problem?

2. You should get a different answer

3. from here for this. [*Points to each path on two-path diagram.* ]

4. **S2**: No no no

5. **S1**: They should be equal?

6. **S2**: They should be equal

7. **S1**: Why should they be equal?

8. This path is longer if you think about it. [*Points to two-part path again*]

9. **S2**: Because force, err, because

10. work is path independent.

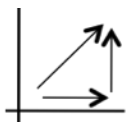

11. **S1**: This path is longer, so it should have, [*Points to two-part path again*]

12. this number should be bigger than

13. **S2**: Work is path independent. If you

14. go from point A to point B,

15. doesn't matter how you get there,

16. it should take the same amount of work.

Lines 1 to 6 contain the main issue of this episode. While S1 thinks there should be different amounts of work done on the small mass along the two different paths, S2 believes the work done should be the same.

S1 provides a justification for his claim in lines 7 and 8 when he challenges S2's same-work assertion. "This path is longer if you think about it." The mathematical definition of work, $W_{A \to B} = \int\limits_{A}^{B} \vec{F} \cdot d\vec{r}$, is essentially "force times distance".  Since the two-part path from A to B is physically (geometrically) longer than the direct route, it seems to follow to S1 that more work must be done along the longer path.

In the language of formal argumentation theory,[35] S1 makes the claim that more work is done along the two-part path, and he offers the data that the two-part path is longer. An unspoken warrant exists that connects his data to his claim: the particular mathematics being used should align with the physical systems under study. The goodness-of-fit between the math at hand and the physical system attests to the validity of one's conclusions. The work formula seems to say "force times distance" to S1.  The two-part path has more "distance", and S1 thus draws justification for his answer.

S1's warrant thus suggests he is framing his activity as physical mapping. His use of a diagram in lines 1 to 3 and 7 to 8 supports this characterization. He gestures to the different paths as he points out that the two-part one is physically longer. Use of a diagram as intermediary between the physical situation and the mathematics is a commonly observed indicator of a physical mapping framing.

S2 not only has a different answer than S1, but he is also framing his use of mathematics in a different way. S2 claims that the work done on the small mass should be the same along the two paths "because work is path independent" (lines 9 and 10). His data is a familiar mantra (though he omits mentioning how this statement is only valid for conservative forces like gravity). The unspoken warrant that S2 is relying on concerns the common use of rules and definitions in math and physics: sometimes previous results are simply taken as givens for speed and convenience. S2 is framing his math use as invoking authority.

After hearing S2's counterargument, S1 repeats himself. In lines 11 and 12, he restates his longer-path justification and again points to the relevant features of the diagram they had previously drawn on the board. S2 responds by restating "work is path independent" in line 13 and again, slightly differently, in lines 14 to 16.

The most important observation in this first clip is that S1 and S2 are disagreeing over much more than merely the answer itself. Explicitly, they are debating whether or not more work is done along the longer path. Implicitly, they are arguing over the most useful way to frame their present use of mathematics. S1 never explicitly says "Please respond to my claim in a way that maps our math to some detail of the physical situation I may have overlooked". His phrasing and gesturing in his initial argument (lines 7 and 8) and beyond (lines 11 and 12) implies this framing request, though.

When S2 responds with his rule citation, he is not merely arguing for a different answer. He is pushing for a different type of warrant for judging the validity of a given answer. S2's invoking an authority framing may have even prevented him from really hearing what S1 was saying. S1's framing request may have passed by S2 unnoticed because he was too caught up in the subset of all his math resources that his invoking authority framing had activated within his mind. At any rate, S2 responds in lines 9 and 10 with a different type of justification than what S1 was expecting.

When S1 repeats himself in lines 11 and 12, he is implicitly repeating his bid for a physical mapping framing. One can imagine a situation when S2's invoking authority justification would simply be accepted without incident, but here it did not align with S1's present framing. S2 does not respond to this reframing request and repeats his answer as he remains in invoking authority.

There is thus an intense framing argument going on under the surface of this debate. Sensing that he is not making any headway in the framing battle, S1 now moves to shift both himself and S2 into a third framing.

### 3. A temporary agreement on a third framing

S1 now makes a move toward a third way of addressing the mathematics at hand. S2 accepts for a time.

17. **S1**: OK, that's assuming Pythagorean

18. Theorem and everything else add[s].

19. Well, OK, well is this— what was the

20. answer to this right here? [*Points to*

$$\int_{\sqrt{2}}^{3\sqrt{2}} \frac{1}{r^2}\, dr = \int_1^3 \frac{1}{y^2+9}\, dy + \int_1^3 \frac{1}{x^2+1}\, dx\cdot]$$

21. What was that answer?

22. **S2**: Yeah, solve each integral numerically.

23. **S1**: Yeah, what was that answer?

24. **S3**: Each individual one?

25. **S1**: Yeah, what was

26. **S3**: OK, let me, uhh    [*S3 starts typing into Mathematica*]

27. **S1**: Cause path two is longer than path one, so

28. **S2**: May I, for a minute? [*S2 writes on a small corner of the blackboard, but never speaks about what he writes.*]

29. **S1**: and path one was this.

30. **S2**: Gimme this, I wanna think about something.

31. **S1**: Just add those up, tell me the number for this [*Points to integrals again*]

32. and I'll compare it to the number of

33. **S3**: OK, the y-one is point one five.

34. **S1**: I, just give me the, just sum those up.

35. I just want the whole total.

36. I just want this total quantity there,

37. just the total answer. [*Points to integrals again*]

38. **S2**: Oh, it was point four-

39. **S3**: No, that's the other one [direct path].

40. **S1**: you gave it to me before, I just didn't write it down.

41. **S3**: Oh I see, point, what, point six one eight

42. **S1**: See, point six one eight, which is what I said,

43. the work done here should be larger

44. than the work done here 'cause the path    [*Points to two-path diagram*]

45. **S2**: No, no no, no no no

46. **S3**: the path where the x is changing

47. **S2**: Work is path independent.

48. **S1**: How is it path independent?

49. **S2**: by definition

50. **S3**: Somebody apparently proved this before we did.

S1 attempts to reframe the discussion in lines 19 to 21. He points to the integrals they've written and asks, "Well, OK…what was the answer to this right here? What was that answer?" He is calling for someone to evaluate each of their expressions for the work so that he can compare the numeric results. This argument relies on another kind of warrant. Mathematics provides one with a standardized, self-consistent set of manipulations and transformations. Performing a calculation or having a computer do it for you according to these rules will give a valid, trustable result. S1 is moving to reframe their math use as calculation.

Even though S1 doesn't explicitly detail the new warrant he is proposing, S2 quickly zeroes in on it. He immediately responds, "yeah, solve each integral numerically" (line 22). Compare this successful, fluid epistemic frame negotiation with the struggle of the previous snippet. Lines 1 to 16 had S1 pushing for physical mapping while S2 lobbied for invoking authority. Both stuck to their positions, resulting in an inefficient conversation. Neither was accepting what the other was trying to say. Lines 19 to 22 have S1 and S2 agreeing, for the moment, on what type of mathematical justification should count.

The calculation framing negotiated, lines 23 to 41 are mostly about S1 directing S3 to input the proper expressions into Mathematica, a common software calculator package. They finish with Mathematica in line 41. It turns out that the radial path integral, $\int_{\sqrt{2}}^{3\sqrt{2}} \frac{1}{r^2} dr$, is equal to 0.47 while the two-part path integrals, $\int_1^3 \frac{1}{y^2+9} dy + \int_1^3 \frac{1}{x^2+1} dx$, evaluate to 0.618. S2 was correct back in lines 1 to 16. The same amount of work should be done along the two paths. While the radial integral is correct as written (within a negative sign), they have neglected the cosine term from the dot product $\vec{F} \cdot d\vec{r}$ in the two-part path integrals.

S1 takes the result of their calculation argument to support his earlier physical mapping framing. "See, point six one eight, which is what I said, the work done here should be larger than the work done here 'cause the path…'" (lines 42 to 44). This move is quite impressive. Here, S1 is using his calculation framing as a subroutine of sorts. He is nesting his computation within a larger scheme of supporting his physical mapping argument of longer-path-means-more-work.

S1 gives another hint that the physical mapping framing has not completely decayed while they are calculating. In the midst of the Mathematica work, he tosses in "cause path two is longer than path one" (line 27). This example illustrates the "hybrid" point made in the interrater reliability section of this paper. Physics students' thinking is simply not always compartmentalized. The four framings only represent general clusters of reasoning. That S1 tosses in a still-active piece from his previous physical mapping into the calculation is neither an anomaly of thought nor a failure of this paper's framework. A likely mark of expertise in physics is a fluid movement among framings. Indeed, this problem was set up and assigned for the very purpose of encouraging students to look for coherency among various framings like S1 is doing here.

Earlier, we claimed that less than 50% of a random episode of student thinking could be cleanly coded as an elemental form of calculation, physical mapping, invoking authority, or math consistency. Still, we claimed that about 90% of a transcript could be seen as a molecular combination of overlapping bits of them. Lines 19 to 41 are an example that is mostly calculation but is fuzzed somewhat by physical mapping.

S2 responds in a familiar way to S1's recall of physical mapping in lines 42 to 44: "See, point six one eight, which is what I said, the work done here should be larger than the work done here 'cause the path". S2 returns to invoking authority to justify his equal-work assertion in lines 45 and 47. "No, no no, no no no…work is path independent". When S1 presses him for more detail, "how is it path independent?" (line 48), S2 and S3 respond "by definition" (line 49) and "somebody apparently proved this before we did" (line 50).

#### 4. An even stronger bid for physical mapping

The replies of S2 and S3 in lines 49 and 50 do not contain the type of justification S1 seeks. The next block of transcript begins with S1 making another strong bid for physical mapping.

51. **S1**: OK, I don't understand the concept then,

52. because you're saying it's path independent.

53. **S2**: I'm saying, if you're at the bottom of a hill

54. **S1**: all right

55. **S2**: and you want to drive to the top of the hill

56. **S1**: right

57. **S2**: and there's a road that goes like this,

58. a road that goes like this, and a road that's like this,

[*Draws* 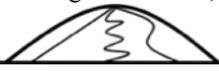 ]

59. it takes the same amount of energy to get

60. from the bottom to the top.

61. It doesn't matter which one you take.

62. **S1**: OK, then you tell me this then;

63. work is force times distance, right?

64. **S2**: It's the integral of f-dr...f-dr, yeah.

65. **S1**: So if you're going this r, and

[*Draws* 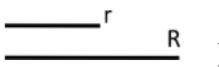 ]

66. you're going this R, which one has more work?

67. **S2**: If there's constant force?

68. **S1**: Constant force on each on

69. **S2**: This one if it's the same force.   [*Points to long "R" path*]

70. **S1**: OK, now the same force is acting on that

71. **S2**: No. No no. Because this one [radial] has

72. direct force the whole time.

73. See, there's lesser force. [*Gestures at two-path diagram*]

74. **S1**: OK

75. **S2**: in each one of these [two part path]

76. **S1**: OK. All right

77. **S2**: your forces are

78. **S1**: I see what you mean.  I see what you mean.

79. Here we're taking

80. **S2**: Here we're supposed

81. to be compensating for that

82. **S1**: We're just taking the x component   [*Gestures at two-path diagram*]

83. of the force here, and the y component

84. of the force there. You're probably right.

85. You've probably been right the whole time

86. Are we thinking about that correctly then?

87. I agree with what you're saying.

S1 begins this last transcript chunk with another bid to frame their math use as physical mapping.  "I don't understand the concept then, because you're saying it's path independent" (lines 51 and 52).

S2 responds to this newest bid with an interesting hybrid of his own.  He is still quoting "work is path independent" but he now couches that rule in terms of a physical situation. He draws a picture of various paths up a hill and asserts "it takes the same amount of energy to get from the bottom to the top. It doesn't matter which one you take" (lines 53 to 61).

S2's latest response still partly reflects an invoking authority framing because it offers no physical mechanism for why the work done by gravity should be the same along any of the paths up the hill. Technically, your car would burn more gasoline along the curviest path, but S2 doesn't acknowledge this point and may not have even considered it in light of the inertia invoking authority is exhibiting in his thought. Perhaps S2 has a more detailed physical mechanism in his mind, but he doesn't articulate it here.

Nonetheless, S1 recognizes a glimmer of the type of justification he seeks in S2's latest argument. S1 presses further on the longer-path issue. "OK, then you tell me this then; work is force times distance, right? … So if you're going this *r*, and you're going this *R*, which one has more work?" (lines 62 to 66) This question is S1's most explicit call yet for a physical mapping framing. He closely juxtaposes a mathematical point (work is force times distance) and a diagram-aided observation of a longer path (his *r* and *R* picture).

This reframing bid tips S2. His hint of a physical mapping framing in lines 57 to 61 asserts itself, putting him in a much better position to understand S1's argument. For the first time in this conversation, S2 explicitly addresses a physical detail relevant to the physical mapping S1 is attempting: "if there's constant force?" (line 67) S1 quickly affirms that assumption and S2 correctly concludes that more work would be done on the long "*R*" path. When S1 quickly moves from this hypothetical *r* and *R* case back to the homework problem (line 70), S2 immediately points out the inconsistency. "No. No no. Because this one [radial] has a direct force the whole time. See, there's lesser force … in each one of these [two-part path]…here we're supposed to be compensating for that" (lines 71 to 75 and 80 to 81). S2 gestures to the problem's diagram during this physical mapping. The gravitational force vector and the displacement vector are (anti-) parallel for the radial path, hence you need to consider the full magnitude of the gravitational force in calculating the work done along that path. These two vectors do not align perfectly along the two-part path, hence you only consider a component of the force there.

S1 quickly accepts and confirms this argument (lines 78 to 87), which is the first fully articulated physical mapping argument S2 had offered during this conversation. His quick comprehension and acceptance occur because S2 has now framed their problem solving in the way S1 has. S1 was mentally ready to accept such an argument.

S2's reluctance to adopt a physical mapping framing implies an activation failure, not lack of sophistication or naivety. His reluctance was certainly not due to simple inability. He was, after all, the one who actually wrote the integrals (which do not contain the necessary cosine factors but, according to S2, were meant to reflect the "lesser force" idea) in the minutes leading up to the presented transcript. S2 quickly generated a physical mapping argument once he framed the discussion as physical mapping, i.e. once he activated the relevant subset of his mental resources.

### 5. Summary of the first case study

This case study illustrates how epistemological framing negotiation and communication can be a powerful dynamic in physics students' work. S1 and S2 disagreed over much more than whether the gravitational work done was independent of path. Their disagreement over what type of justification was appropriate drove this conversation. Much of this debate was implicit. S1 never came out and said, for example, "please respond to me in a way that points out some detail of the physical situation that I have not mapped correctly to the mathematics we're using." The epistemological framing analysis presented in this dissertation offers a way of making this implicit conversation-dynamic explicit to teachers and physics education researchers.

In this case study these epistemological framings exhibiting had considerable inertia. S2 remained in invoking authority despite several prods. S1's commitment to physical mapping allowed those prods to keep happening. The next case study shows a student shifting frames much more readily.

## B. Case study 2: Framings can be flexible

This next case study also has two physics students trying to agree on the best way to frame the math use at hand. Like S1 in the last example, S4 will make several framing bids. S5 responds to these bids more readily than S2 did, illustrating how epistemological framing can be a relatively labile process as well.

### 1. The question

The two students in this episode are enrolled in a second semester undergraduate quantum mechanics class. Like the students in the previous episode, they are meeting outside of class to work on that week's homework assignment. The case study begins with the students part way through problem 6.32, part b, in Griffiths's *Introduction to Quantum Mechanics*, a common undergraduate textbook.[50] That problem deals with the Feynman-Hellmann theorem,

$$\frac{\partial E_n}{\partial \lambda} = \left\langle \psi_n \left| \frac{\partial H}{\partial \lambda} \right| \psi_n \right\rangle,$$ which relates the partial derivative of an energy eigenvalue with respect to any parameter $\lambda$ to the expectation value of the same partial derivative of the Hamiltonian. The problem tells them to consider the one-dimensional harmonic oscillator, for which the Hamiltonian is $H = \frac{-\hbar^2}{2m}\frac{\partial^2}{\partial x^2} + \frac{1}{2}m\omega^2 x^2$ and the $n^{\text{th}}$ eigenvalue is $E_n = \hbar\omega\left(n + \frac{1}{2}\right)$. They are asked to set $\lambda$ equal to $\omega$, $\hbar$, and $m$ (the angular frequency of the oscillator, Planck's constant, and the mass of the oscillator, respectively) and to use the Feynman-Hellmann theorem to get expressions for the oscillator's kinetic and potential energy expectation values.

We begin with S4 noticing an oddity. When she sets $\lambda = \hbar$, the Feynman-Hellmann theorem requires her to consider $\frac{\partial}{\partial \hbar}$. How does one deal with a partial derivative with respect to a constant?

### 2. A framing clash and a quick shift

The two students are seated at a table throughout this discussion. They do not gesture towards any diagrams or equations in a shared space.

1. **S4**: If we figure this out, hopefully it'll make

2. the other ones easier. When you say something's

3. a function of a certain parameter, doesn't that mean

4. that as you change that parameter, the function changes?

5. **S5**: mmm-hmm

6. **S4**: OK, so I can change omega, but I can't change h-bar.

7. **S5**: Sure you can.

8. **S4**: I can?

9. **S5**: You can make it whatever you want it to be.

10. **S4**: But

11. **S5**: It's a constant in real life, but it's a funct-, it's,

12. it appears in the function and you're welcome to change its value.

13. **S4**: But then it doesn't mean anything.

14. **S5**: Sure it does.  Apparently it means

15. the expectation value of [kinetic energy].

16. **S4**: You don't really know what you're talking about.

17. **S5**: Look, all it is, is you're gonna take the derivative with respect to

18. **S4**: Yeah, I understand what they want me to do here.

19. **S5**: They're just applying the theorem.

S4 begins this passage with a concise check on what a derivative entails.  "When you say something's a function of a certain parameter, doesn't that mean that as you change that parameter, the function changes?" (lines 2 to 4).  Upon S5's affirmation, S4 points out a mismatch of this mathematical point with a physical reality.  The parameter $\hbar$ is a physical constant.  Taking a partial derivative with respect to $\hbar$ would imply that Planck's constant can vary.  S4 is framing her use of mathematics as physical mapping.  Her warrant for not accepting the $\frac{\partial}{\partial \hbar}$ operation focuses on how valid uses of math in physics class tend to align with physical reality.

S5 initially responds to S4's concern by asserting a rule.  The warrant for his counterargument concerns the practical, common use of statements and previous results without explicit justification.  "Sure you can [change $\hbar$]" he says.  "You can make it whatever you want it to be" (lines 7 and 9).  In so responding, S5 is lobbying for an invoking authority framing.  He is suggesting S4 set aside her physically motivated objections and instead judge the validity of $\frac{\partial}{\partial \hbar}$ according to his confidence in his assertions.

Much like the two students in the gravitational work example, S4 and S5 are arguing over something much deeper than whether or not one is allowed to take a partial derivative with respect to $\hbar$.  They are disagreeing over what would be appropriate grounds for accepting or rejecting such an operation.

S4 does not accept S5's bid for invoking authority.  Upon her first protest in line 10, S5 quickly admits "it's a constant in real life" (line 11) but sticks to his invoking authority framing.  "It appears in the function and you're welcome to change its value" (lines 11 and 12).

S4 protests again; "But then it doesn't mean anything" (line 13).  Such a statement's full interpretation relies on acknowledging S4's physical mapping framing.  In some framings, S4's statement is patently false.  The operation $\frac{\partial H}{\partial \hbar}$ can "mean" plenty.  For example, formally carrying out the operation on the Hamiltonian operator would produce the operator $\frac{-\hbar}{m}\frac{\partial^2}{\partial x^2}$.  Developing both the calculus machinery and the abstract interpretation of such an operation was the crowning achievement of Newton and Leibniz's mathematical studies.  S5 retains his invoking authority framing and quickly responds with another "meaning" of $\frac{\partial H}{\partial \hbar}$.  Quoting from the textbook's statement of the homework problem, "Sure it [means something].  Apparently it means the expectation value of [kinetic energy]" (lines 14 and 15).  Recall the question had told them to set $\lambda = \hbar$ in the Feynman-Hellmann theorem, $\frac{\partial E_n}{\partial \lambda} = \langle \psi_n | \frac{\partial H}{\partial \lambda} | \psi_n \rangle$, and hence obtain an expression for the expectation value of kinetic energy.  S5 is thus relying on the authority of the text's question for his interpretation of $\frac{\partial H}{\partial \hbar}$.  Only by acknowledging S4's current physical mapping framing can we place her claim in the proper context.  If one's warrant for judging an operation like $\frac{\partial H}{\partial \hbar}$ concerns the alignment of the mathematics with a physical reality, then yes, that operation can be said not to "mean" much of anything.  In the real physical world Planck's constant has a particular value and does not vary.

S4 objects to S5's arguments again in line 16.  "You don't really know what you're talking about."  This perturbation was sufficiently strong to cause S5 to reframe his attempt to justify $\frac{\partial}{\partial \hbar}$.  He says "look, all it is, is you're gonna take the derivative with respect to" (line 17) before getting cut off by S4.  Coupled with his next statement in line 19, "they're just applying the theorem," these statements can be seen as an attempt to reframe his thinking as calculation.  S5 is suggesting they go ahead and use their calculus machinery to take the partial derivative.  As long as they stay true to the rules of calculus, they should be able to trust whatever result appears.

S4 acknowledges this attempt to reframe their work as calculation.  "Yeah, I understand what they want me to do here" (line 18).  Lines 17 to 19 nicely illustrate how efficient this implicit epistemic frame negotiation can be. These lines didn't even take five seconds to speak. In those five seconds, S5 made a call for using a different set of warrants. S4 heard that call and her brain quickly activated some of the procedures and techniques that would be associated with such a framing, as evidenced by "yeah, I understand what they want me to do here" (line 18). S5, just as quickly, acknowledges S4's acknowledgment of his reframing suggestion with his "they're just applying the theorem" (line 19).

### 3. Another quick shift, this time to a shared physical mapping framing

S4 still insists on a justification more in line with her physical mapping framing. She begins the next chunk of transcript with another reframing objection. S5 responds by nimbly dropping his calculation framing and adopting physical mapping himself.

20. **S4**: But I don't understand how you can take the derivative

21. with respect to a constant.

22. **S5**: Because if you change the constant then the function will change.

23. **S4**: But then it's not, it's not physics.

24. **S5**: So? Actually it is, 'cause, you know,

25. a lot of constants aren't completely determined.

26. **S4**: There's still only one value for it, that's what a constant is.

27. **S5**: The Hubble constant changes. The Hubble constant changes

28. as we improve our understanding of the rate of expansion of the universe,

29. and we use the Hubble constant in equations.

30. **S4**: But there's only one, right, there's only one constant. It does not vary.

31. **S5**: Yeah, but the value's changing as we approach the correct answer.

32. **S4**: It's just gonna get fixed. That's not, that's not helping us with the derivative.

33. **S5**: You can always take a derivative with respect to anything.

34. **S4**: But if you take it with respect to a constant, you'll get zero.

35. **S5**: Not if the constant itself appears in it.

36. The derivative tells you if you change whatever

37. you're taking the derivative with respect to how will the function change?

S4 begins this block of transcript by repeating her discomfort with $\frac{\partial}{\partial \hbar}$ (lines 20 and 21). S5 responds with "because if you change the constant then the function will change" (line 22). This statement does not clearly align with only one of this paper's common framings. Its ambiguity comes in large part from its isolation. Perhaps it was a prelude to a calculation explanation, or perhaps S5 was preparing to use some sort of Math Consistency warrant as he related this $\frac{\partial}{\partial \hbar}$ issue to a more familiar Calculus 101 example. S5's thought could have evolved this way or that, but one cannot assume line 22, by itself, was necessarily the tip of an implicit iceberg of coherence.

S4's next objection, "but then it's not, it's not physics," (line 23) leads S5 to start explicitly searching for an example of a physical constant that varies. In undertaking such a search, S5 has adopted the framing S4 has been pushing. Valid use of math in physics class should align with physical reality. S5 hopes that by finding an example of a varying physical constant he can convince S4 that it is permissible to take a derivative with respect to Planck's constant. S5 frames his activity as physical mapping starting in line 24.

S5 invokes the analogy of the Hubble constant in lines 24 to 31. The Hubble constant is connected to the rate of expansion of the universe. S5 points out that the value of the Hubble constant quoted by scientists has changed over the past half a century as our measurement techniques have improved. He argues that the Hubble constant, variable as it seems, is an important part of many physics equations. By extension, it should be permissible to consider a varying Planck's constant.

S4 offers a much richer response to S5's Hubble constant argument than she has to any of his other attempts in this episode. Up to this point, she had been simply shooting down S5 with comments like "but then it doesn't mean anything" (line 13), "you don't really know what you're talking about" (line 16), and "but then it's not, it's not physics" (line 23). S5's Hubble constant argument marked the first time he adopted S4's warrant concerning the alignment of math and physics, i.e. the first time he and S4 shared a common epistemological framing.

This shared epistemological framing helps S4 engage with S5's chosen example in lines 26 to 32, and she points out that he's confusing a measurement variance with an actual physical variance. Sure, she says, our quoted value for the Hubble constant has shifted as our measurements improve, but, presumably, our measurements are tending towards a fixed value. The Hubble constant itself, she says, isn't changing. "That's not helping us with the derivative" (line 32).

This counterargument causes S5 to reframe the situation once again as he turns to a different type of justification. He quotes a rule again in line 33. "You can always take a derivative with respect to anything." S4 misspeaks when she replies. "But if you take it with respect to a constant, you'll get zero" (line 34). This statement seems to confuse her earlier correct interpretation of $\frac{\partial}{\partial \hbar}$ (as in lines 2 to 6) with the Calculus 101 mantra "the derivative of a constant is zero", i.e. $\frac{\partial \hbar}{\partial x} = 0$. S4 responds to this misstatement in lines 35 to 37.

### 4. A final frame shift

The final block of transcript from this episode follows S5's quick correction. It begins with S4 objecting yet again and S5 trying out yet another framing.

38. **S4**: So I don't understand how you can change a constant.

39. **S5**: You pretend like it's not a constant.

40. It's just like when you take partial derivatives with respect to,

41. like variables in a function of multivariables.

42. You pretend that the variables are constant.

43. **S4**: Yeah, I don't have a problem with that.

44. **S5**: You're going the other way now.

45. You're pretending a constant is a variable. Who cares?

46. **S4**: It doesn't make sense to me.

47. **S5**: You can easily change a variable—it's not supposed to, I don't think.

48. **S4**: OK, then I believe-

49. **S5**: I don't think, I don't think there's supposed to be

50. any great meaning behind why we get the change h-bar.

51. I think it just-they're like oh look, if you do it

52. and you take its derivative and you use this equation,

53. then all of a sudden you get some expectation of [kinetic energy],

54. and you say whooptie-freekin-do.

S5 responds to S4's latest objection in line 38 via a math consistency framing. His newest argument relies on a warrant he hasn't yet tried: mathematics is a self-consistent field of knowledge, so a valid mathematical argument is one that fits in logically with other mathematical ideas.

S5 makes a common move for a math consistency framing. He draws an analogy in lines 39 to 45. In order to take a derivative with respect to $\hbar$, one has to "pretend" that the constant is a variable. S5 points out that taking a standard partial derivative with respect to one of the variables of a multivariable function involves "pretending" the other variables are constants. Their $\frac{\partial}{\partial \hbar}$ case, he argues, is "just like" that analogous example, except "you're going the other way now. You're pretending a constant is a variable."

In contrast to her more extended counterargument in the Hubble constant case, S4 rejects this present argument much more coarsely. "It doesn't make sense to me" (line 46). S5 has once again framed their work differently than S4's physical mapping. A plausible explanation is that each student's mind has activated a sufficiently different subset of their available mathematical resources, and that restricts the depth of their communication and interaction.

When S5 responds "it's not supposed to [make sense], I don't think" in line 47, he is explicitly addressing S4's physical mapping framing for the first time. While he had been responsive to her objections throughout this episode, he now argues with her epistemological framing directly. He states that he doesn't think an explanation of the type S4 seeks exists. S4 is possibly about to acknowledge inappropriateness of the physical mapping stance when she replies "OK, then I believe-" (line 48), but she gets cut off. S5 then elaborates a hybrid of calculation and invoking authority that he sees as most appropriate in lines 49 to 54. Mechanically take the derivative with respect to $\hbar$, following the familiar calculation algorithms, and then trust the Feynman-Hellmann theorem to relate this derivative to the oscillator's kinetic energy.

### 5. Summary of the second case study

This case study illustrates how epistemological framing can be a relatively flexible process. The entire episode is essentially many iterations of S4 objecting and S5 saying, "Well, all right, how about this other type of explanation?" S4's objections serve as perturbations to S5's mental state. Many of them are of sufficient strength (or occur after he has reached a respectable closure point of his previous argument) to lead him to reframe his thinking. Each reframing results in S5 adopting a different type of warrant for judging the validity of his mathematical claim, that one should accept the operation $\dfrac{\partial}{\partial \hbar}$ as legitimate within physics, despite the constancy of $\hbar$.

This $\dfrac{\partial}{\partial \hbar}$ issue is a relatively difficult one. Ordinarily, a physical mapping frame is quite valuable in physics. Helping students understand the physical referents or their math is a common, if sometimes difficult, goal of many physics classes. Here, S4 and S5 are being asked to do something even more subtle and difficult: consider an imaginary world, one where $\hbar$ can vary, and see if the mathematics in this imaginary world can inform the real one. That S4 and S5 were willing to engage in an exploration of how to frame this $\dfrac{\partial}{\partial \hbar}$ issue is commendable, even if the episode ends without an especially satisfying consensus.

# VI. CONCLUSION AND IMPLICATIONS

## A. Summary

In this paper we have argued that analyzing student problem solving from the point of view of the kind of warrants (or epistemological resources) they choose to use gives insight into the way the student is framing the mathematical task at hand. From a large number of ethnographic observations of students in upper division physics classes we selected situations in which students were taking contrasting views on the approach to be used. From this salient data we created a classification of warrants that we believe indicate the students' epistemological framing of the task.

The two case studies both align and contrast with each other. Both demonstrate how epistemological framing dynamics can drive a conversation. The first illustrates that these framings can have significant inertia (as with S2), while the second shows that they can be relatively flexible (as with S5). Both demonstrate that using a warrant analysis gives us additional insight into the students' reasoning, concerns, and possible errors.

The students in each case study disagree over much more than an answer. They each frame their activity differently and hence try to apply a different type of warrant to judge the validity of their claims. The students exert various pushes and pulls on each other as they try to negotiate a common epistemological framing. Vary rarely are these framing bids explicit. Nonetheless, these framing debates underlie the speech in both of our case studies. When a common framing is established, the conversation tends to be richer and more efficient. The warrant analysis presented here is meant as a useful tool for finding explicit evidence of what is usually an implicit cognitive process.

## B. Implications for instruction

Our analysis has important implications for teachers as well as for researchers. First, being aware of framing can help an instructor be aware of when he and his students are not communicating – when they are "not on the same page." Second, being aware of framing can lead an instructor to understand the value of hybrid and flexible framing and lead to her evaluating student progress in a fashion that is both subtler and more productive. We elaborate briefly on each of these points and speculate on how an instructor might respond to these issues effectively.

### *1. Being aware of framing can reveal failures in communication*

Framing differences like those in the case studies here, and the miscommunications that accompany them, could certainly occur between instructor and student as well as between students. Both the instructor and students will naturally frame what occurs during a lesson, but there is no guarantee they will frame each part of the lesson in the same way. A teacher may calculate for a while and then want to make a point about how an equation matches a physical expectation. The teacher may even offer a signal that she's switching approaches, but perhaps that signal isn't sufficient to tip the students. They may merely try to interpret her physical mapping comments through a calculation lens, or even reject the physical mapping reframing as irrelevant and stop paying attention. Perhaps a professor gives an extended math consistency discussion, carefully explaining how the math at hand is analogous to a more familiar math idea. Maybe her students are framing his discussion as invoking authority and instead hear a series of math facts to be accepted on faith.

We conjecture that there are (at least) two ways to combat such teacher/student framing misunderstandings. The first is for a teacher to simply exaggerate her framing cues. If the situation calls for conven-iently quoting a rule, spend a little extra time explaining your reasons for doing so. If it's obvious to you, as a teacher, that a physical mapping discussion is in order, make that (and your reasons for believing so) more explicit to your class. More explicit framing cues might lessen the probability of miscommunication due to a framing mismatch.

A second antidote to teacher/student framing mismatches is for the teacher to gather more evidence, in real time, of her students' framing. In a traditional lecture, information tends to only flow from the professor to the students. Such a lecturer will have scant evidence available for how her students are framing her lesson. Asking questions that have simple phrase-like answers may give a teacher evidence of the simple correctness or incorrectness of the class's answers, but is only of marginal help for deducing the students' epistemological framing. Engaging one's students in extended discussions during class is the best way to get valuable framing evidence. Asking open-ended questions that give students a wide range of possible responses will require them to explain their reasoning to a much greater depth. As they explain their justifications for their claims, their framing will become much more apparent to the teacher. Framing mismatches will become much easier to diagnose in real time.

### 2. Being aware of framing can help an instructor better evaluate the progress of a student

We conjecture that expertise in physics problem solving involves the ability to blend different epistemological framings and to flip quickly from one framing to another in response to snags and difficulties. The clustering of skills into frames can be an efficient way to proceed; if one can quickly recognize the tools needed to solve a problem, the problem can be solved without spending time hunting through a large number of possibilities. But when problems or inconsistencies arise, it can be more effective to explore a wider search space of possibilities.

If an instructor is aware of the fact that students may "get stuck" in a framing that limits their access to tools and knowledge they may not only possess but be good at, the instructor will have a better understanding of the true nature of the difficulty the students may be experiencing. That instructor will be less likely to "write off" his students as incompetent and more likely to try prodding them into a different framing. Developing homework questions comes to be seen as creating tasks of sufficient richness and complexity to help students develop these frame-juggling skills on their own. This issue is discussed in more detail in reference 1b, Chapter 7.

## C. Implications for future research

This analysis opens possibilities for significant research efforts by illuminating a dimension of student performance that is rarely considered as a component of "student difficulties" but that potentially plays a critical and controlling role for many students. Much more work is needed, both in improving the methodology of identifying student framing and in explicating the role framing difficulties play in the typical classroom.

## ACKNOWLEDGMENT

We gratefully acknowledge the support and suggestions of many members of the University of Maryland Physics Education Research Group and of visitors to the group including David Hammer, Andrew Elby, Rachel Scherr, Rosemary Russ, Ayush Gupta, Brian Frank, Saalih Allie, and Steve Kanim. This material is based upon work supported by the US National Science Foundation under Awards No. REC 04-4 0113, DUE 05-24987, and a Graduate Research Fellowship. Any opinions, findings, and conclusions or recommendations expressed in this publication are those of the author(s) and do not necessarily reflect the views of the National Science Foundation.